\documentclass[%
 reprint,
 amsmath,amssymb,
 aps,
floatfix,
]{revtex4-2}

\usepackage{graphicx}% Include figure files
\usepackage{dcolumn}% Align table columns on decimal point
\usepackage{xcolor}
\usepackage{subfigure}
\usepackage{amsmath}
\usepackage{amssymb} 
\usepackage{mathrsfs}
\usepackage{color}
\usepackage{multirow}
\usepackage[normalem]{ulem}
\usepackage{bm}% bold math

\usepackage{array}
\usepackage{placeins}
\usepackage{physics}
\begin{document}

%\preprint{APS/123-QED}

\title{Steady granular discharge across aperture shapes}

\author{Kevin Yifan Zhao$^{1}$}\thanks{These authors contributed equally to this work.} 
\author{Ram Sudhir Sharma$^{2}$}\thanks{These authors contributed equally to this work.}
\author{Alexandre Leonelli$^{2}$}
\author{Eckart Meiburg$^{2}$} 
\author{Alban Sauret$^{3,4}$}\email{asauret@umd.edu} 

\affiliation{$^{1}$ Department of Physics, University of California, Santa Barbara, CA 93106, USA\\
$^{2}$ Department of Mechanical Engineering, University of California, Santa Barbara, CA 93106, USA\\
$^{3}$ Department of Mechanical Engineering, University of Maryland, College Park, MD 20742, USA\\
$^{4}$ Department of Chemical and Biomolecular Engineering, University of Maryland, College Park, MD 20742, USA}

\begin{abstract}
We recently proposed a framework for the steady mass flux of particles through circular apertures [R.~S.~Sharma \textit{et al.}, Phys. Rev. Lett. \textbf{136}, 068204 (2026)]. 
Here, we combine experiments and discrete-element simulations of spherical particles discharging through triangular and rectangular apertures to test this
framework beyond circular openings.
Despite changes in aperture shape, we observe a similar free-fall flux, with a modified relevant transport length.  
Deviations due to confinement reflect packing modifications, which are explained by geometry. 
A boundary layer dilation imposed by the aperture edge is shown to be a mechanism for the decrease of flux across various aperture shapes.  
By integrating the local packing deficit over the opening, we identify a hydraulic diameter $D_h=4A/P$, where $A$ and $P$ are the aperture area and perimeter, respectively, as the leading confinement length. This length also provides a reference scale, $\sqrt{gD_h}$, for the mean outlet velocity when the aperture is much larger than an individual particle.
The same local packing model predicts the field in a lens-shaped aperture without further fitting, while the normalized discharge rates follow a common trend across the tested shapes.
\end{abstract}

%\keywords{Suggested keywords}
\maketitle

%%%%%%%%%%%%%%%%%%%%%%%%%%%%%%%%%%%%%%%%%%%%%%%%%%%%%%%%%%%%%%%%%%%%%%%%%%%%
%%%%%%%%%%%%%%%%%%%%%%%%  SECTION I: INTRODUCTION  %%%%%%%%%%%%%%%%%%%%%%%%
%%%%%%%%%%%%%%%%%%%%%%%%%%%%%%%%%%%%%%%%%%%%%%%%%%%%%%%%%%%%%%%%%%%%%%%%%%%%

\section*{Introduction}

\begin{figure*}
\centering
\includegraphics[width=0.8\textwidth]{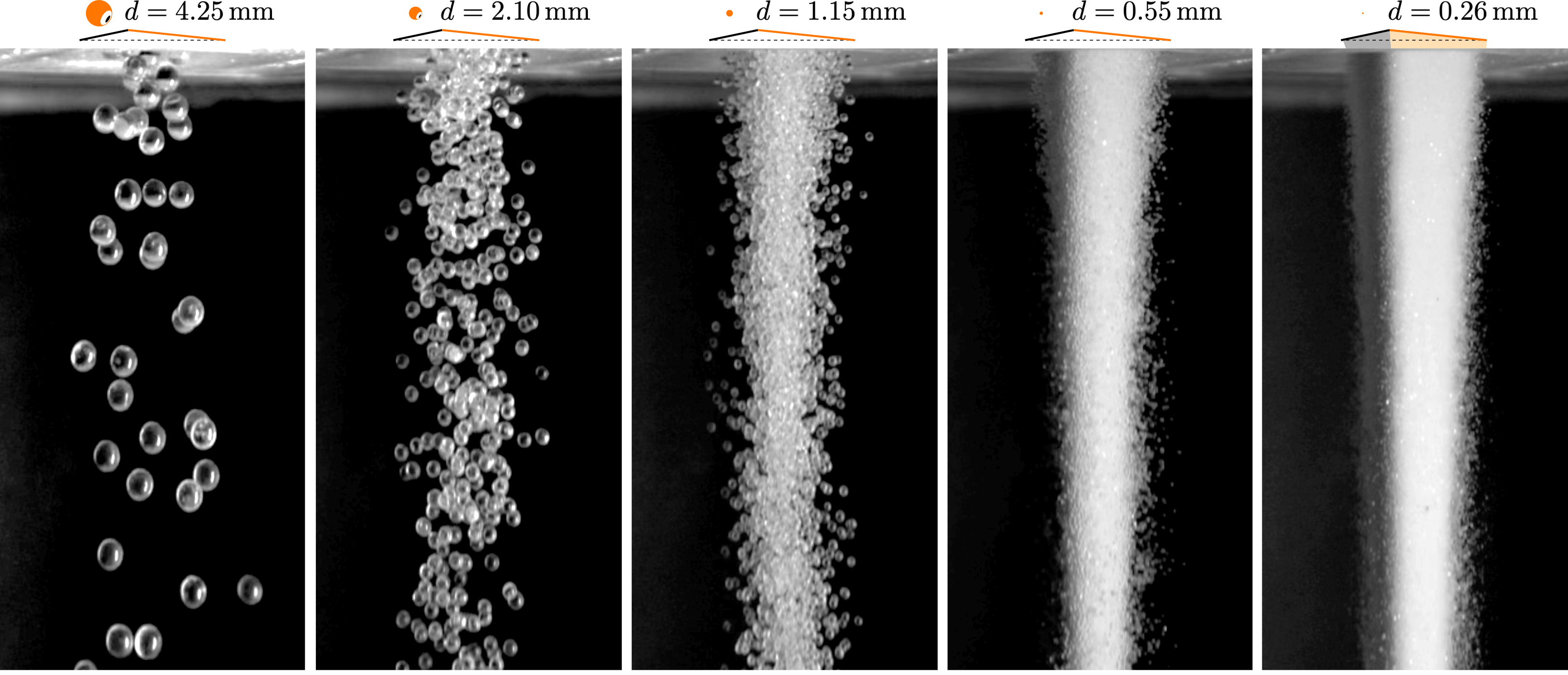}
\caption{Photographs of experiments showing gravity-driven discharge through an equilateral triangle aperture of side $a = 25$ mm for particles of different diameters $d$. From left to right, $d = 4.25,\, 2.10,\, 1.15,\, 0.55,$ and $0.26\,\mathrm{mm}$  ($a/d \simeq 5.9, \, 11.9, \, 21.7, \, 45.5, \, 96.1$, respectively). As the aperture becomes large relative to the particles ($a/d \gg 1$), the discharge becomes dense, and the emerging jet increasingly preserves the triangular cross-section of the opening. The aperture is drawn to provide scale.}
\label{fig:Fig1_imaging}
\end{figure*}

Gravity-driven discharge through a bottleneck is a canonical granular-flow problem with broad relevance to the handling, dispensing, and transport of particulate materials \cite{beverloo1961flow,fowler1959flow}. Its defining feature is that the mass flux remains nearly constant over much of the discharge duration, despite the decreasing weight within the container. Classical investigations by \citet{fowler1959flow} and by \citet{beverloo1961flow} established robust empirical relations between this flux $Q$, the particle size $d$, and the dimensions of the outlet $D$. Although circular apertures later became the standard geometry, these early studies provided a small dataset suggesting that square, rectangular, triangular, and other non-circular openings can be accounted for through a hydraulic diameter. \citet{harmens1963flow} later incorporated circular, polygonal, and oblong apertures through equivalent geometric dimensions, including the hydraulic diameter. Older interpretations also invoke an empty annulus at the aperture edge or a free-fall arch above the opening \cite{brown1970principles}, but without robust measurements of the local packing and velocity, whether the transition near the edges is sharp or gradual is hard to ascertain. More recent three-dimensional studies have also examined how aperture shape affects the discharge rate \cite{wan2018orifice,zaki2019orifice}. These studies showed that the effect of the shape of the aperture can be incorporated into empirical discharge correlations. In the continuum rheology model of \citet{staron2012continuum}, the granular phase is treated as incompressible, so local variations in packing fraction are not resolved. In the energy-balance model
of \citet{darias2020energy}, finite-particle effects at the outlet are incorporated through an effective aperture correction. Although this correction captures their influence on the total discharge rate phenomenologically, it does not predict the local packing deficit near the aperture boundary or its dependence on aperture shape.    
\smallskip

More recent work has shifted attention from global discharge relations toward the local structure of the outlet flow. Measurements in quasi-two-dimensional silos showed that the flux can be decomposed into self-similar velocity and packing-fraction profiles across the opening \cite{janda2012flow, maza2013velocity, zhou2015discharge, gella2017role, bhateja2020velocity}. \citet{aguirre2010pressure} further demonstrated in a horizontal system driven by a conveyor belt that the discharge rate was controlled by the imposed exit velocity rather than by the pressure near the aperture. \citet{rubio2015disentangling} subsequently showed that the classical free-fall arch should not be interpreted as a sharply defined physical structure: the contact and stress-supported material instead crosses continuously into the emerging flow over a region whose extent scales with the outlet size. Together, these studies suggest that the steady flux is most directly understood from the packing and kinematics established locally near the aperture, rather than from detailed mechanical or geometric arguments for the overburden. 
\smallskip

Extensive studies of granular flow through bottlenecks have also shown how constriction size, particle geometry, and arch formation govern flow interruption, particularly when aperture and grain sizes become comparable \cite{zuriguel2005jamming,thomas2013geometry,thomas2015clogging,zuriguel2014review,hafez2021shape}. Clogging also limits transport in non-Brownian particulate suspensions and follows the same physics as in granular flows \cite{vani2024angle,sauret2026clogging}. The quasi-two-dimensional geometry has become an archetypal setup, enabling simulations and experiments to measure local fields, clogging statistics, and stability. However, the 2D simplification removes an important geometric degree of freedom. For a 2D silo, the slit-like outlet is one-dimensional and characterized by its width. In three dimensions, the aperture is two-dimensional: its area, perimeter, corners, and aspect ratio provide distinct geometrical parameters. The question is how these geometric features affect
the local packing and velocity, and hence the steady
discharge rate. Overall, a gap remains between early three-dimensional measurements of shape-dependent discharge and modern field-resolved descriptions of outlet flow. \smallskip

In Ref.~\cite{sharma2026gravity}, we considered the idealized case of spherical particles discharging through a circular aperture. As we approached the limit where the particle is infinitesimal compared to the aperture, the steady flux was well described by a free-fall value: \begin{equation}
Q_{\rm ff} \;=\; A\,\rho_g\,\phi_{\rm ff}\,\sqrt{gD}.
\label{eq:Q_ff_MFI}
\end{equation} The mass flux was separated into velocity and packing contributions, where $\phi_{\rm ff} \simeq 0.55$ is the asymptotic packing fraction at the outlet for unsupported spherical cohesionless particles, and the mean velocity of the exiting particles approached $\sqrt{2gR_0}$. The area of the aperture is $A = \pi R_0^2 = \pi D^2/4$, $\rho_g$ is the density of the grains, and $g$ is the acceleration due to gravity. Departures from this state were associated primarily with boundary-induced dilation, with the packing recovering to the free-fall value far from the boundaries. Near the aperture edge, exponential recovery from zero at the boundary toward the interior packing fraction, with a characteristic lengthscale $\ell\simeq 1 d$, the size of a particle. Capturing the relaxation of both packing and velocity to their asymptotic values, an overall empirical correction with a characteristic length $nd \simeq10d$--$15d$ was provided for the integrated picture over the entire circular aperture. In words, for $D \gg nd$, Eq.~\eqref{eq:Q_ff_MFI} was a good measure for the steady discharge.
\smallskip

In this study, we consider whether this local picture survives when the circular symmetry of the opening is removed. We combine three-dimensional experiments and discrete-element simulations of discharge through rectangular and triangular apertures, supplemented by a predictive test using a lens-shaped opening. The packing deficit remains a local function of distance from the aperture boundary, similar to the circular aperture case \cite{sharma2026gravity}, and its integral over the opening selects the hydraulic diameter $D_h=4A/P$ as the leading confinement length. Independently, the contact-supported packing begins to be released over a distance $D_h/2$, motivating an analogy to the circular case, with the asymptotic velocity scale now $\sqrt{gD_h}$. Altogether, $Q_{\rm ff} = A \, \rho_g \, \phi_{\rm ff} \,  \sqrt{gD_h}$ is shown to capture the steady flux across the tested apertures for $D_h \gg d$. Finite-confinement deviations collapse across the tested aperture shapes, sizes, and particle sizes with the same characteristic length $nd \simeq10d$--$15d$ without needing additional fitting parameters.

%%%%%%%%%%%%%%%%%%%%%%%%%%%%%%%%%%%%%%%%%%%%%%%%%%%%%%%%%%%%%%%%%%%%%%%%
%%%%%%%%%%%%%%%%%%%%%%%%  SECTION II: EXPERIMENTS  %%%%%%%%%%%%%%%%%%%%
%%%%%%%%%%%%%%%%%%%%%%%%%%%%%%%%%%%%%%%%%%%%%%%%%%%%%%%%%%%%%%%%%%%%%%%%

\section*{Experiments}

\begin{figure*}
\centering
\includegraphics[width =\textwidth]{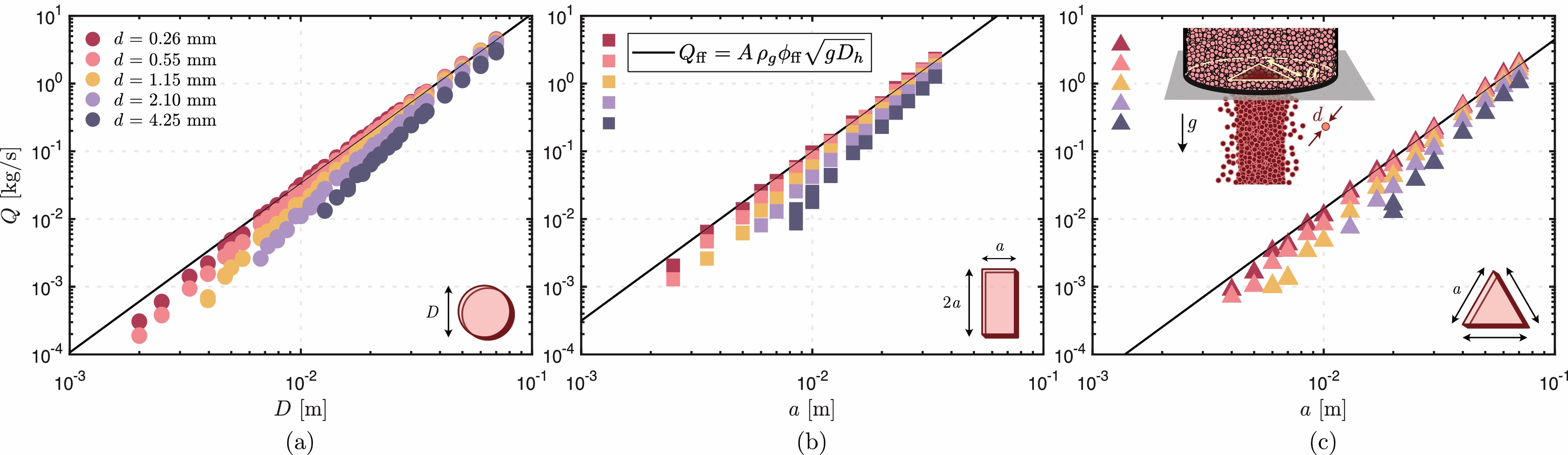}
\caption{Measured steady mass flux $Q$ as a function of the physical aperture size for (a) circular openings of diameter $D$, (b) rectangular openings of sides $a$ and $2a$, and (c) equilateral triangular openings of side $a$. Marker colors denote particle diameter $d$. The solid line in each panel shows the asymptotic free-fall prediction $Q_{\rm ff}$ [Eq.~\eqref{eq:flux_FF_genApp}] where $D_h = 4A/P$ is the hydraulic diameter.
For apertures much larger than the particles, the measured flux approaches the free-fall limit, independent of the opening shape.}
\label{fig:Fig1_appertureFlux_raw}
\end{figure*}

In our experiments, a large flat-bottomed cylinder is first filled with particles and subsequently drained through a central aperture at its base. In each case, spherical glass beads (Sigmund Lindner GmbH) of a single diameter $d = 0.26$, $0.55$, $1.15$, $2.10$ or $4.25 \, \mathrm{mm}$ ($\approx 10\%$ size variation) and density $\rho_g \approx 2500\,\mathrm{kg}\,\mathrm{m}^{-3}$ are used. At rest, inside the cylinder, the grains occupy a volume fraction of $\phi_{\rm cyl} \approx 0.60$. The shape and size of the aperture are varied across experiments. Complete tables of aperture shapes, sizes and particle properties are provided in the Supplemental Material (S.I.). Two cylinders were used. Most experiments were conducted using a stainless steel cylinder of diameter $D_{\rm cyl} = 10.2$ cm, while others used a PMMA cylinder of diameter $D_{\rm cyl} = 20$ cm, such that $A_{\rm cyl} \gtrsim 10 A$, where $A_{\rm cyl}$ and $A$ are the areas of the cylinder and aperture respectively. For the smaller cylinder $3.5$ kgs of mass was initially filled and for the larger cylinder, $5-7$ kgs were filled prior the opening of the aperture.

\smallskip

In Fig.~\ref{fig:Fig1_imaging}, we show images of particle discharge through an equilateral triangular aperture with side length $a = 25\,\mathrm{mm}$ for a range of $d$ in the steady regime. From left to right, we show $a/d$ going from $\mathcal{O}(10^0)$ to $\mathcal{O}(10^2)$. When $a/d \lesssim 4$, we observe clogs, similar to the circular case. For larger but comparable relative sizes $(a/d \gtrsim 4)$, once flows are established, they largely persist uninterrupted and steadily discharge mass. When $a/d \gg 1$, the discharge forms a dense, fluid-like jet with a cross-section that resembles the aperture shape. These observations of discharge through the alternate-shaped apertures follow the same trends as the circular case \cite{sharma2026gravity}. In these cases, the discharge visibly contracts as it falls. For small enough grains, even glass beads can become sensitive to drag forces from the air, and adhesive contact forces may also become relevant \cite{sharma2025models}. We therefore exclude any smaller grains. \smallskip

When the aperture is opened, the discharged grains are collected on a high-precision balance (OHAUS EX6202) that records the accumulated mass $M(t)$ at a sampling rate of $10\,\mathrm{Hz}$ with $\pm 0.01 \, \mathrm{g}$ resolution. The mass flux $Q$ is then obtained as the slope of a linear fit to $M(t)$ over $t \in [T_1, T_2]$, with $T_1$ and $T_2$ defined by $M(T_1) = 0.5\,\mathrm{kg}$ and $M(T_2) = 2.5\,\mathrm{kg}$. This window is chosen to exclude both the initial transient and the final discharge stage, when few grains remain in the column and the free surface deforms in the center above the aperture. Within this window, mass flux through the aperture remains largely steady. Multiple trials are conducted for some combinations of aperture shape, size and grain size. Repeated measurements under identical conditions typically agree within $1\%$, with discrepancies never exceeding $5\%$. The complete set of measurements is provided in a public repository.
\smallskip

We present the measured mass flux $Q$ across the complete range of aperture shapes, sizes, and grain diameters in Figs.~\ref{fig:Fig1_appertureFlux_raw}(a)--(c), where each panel shows the data for a single aperture shape: (a) circular, reproduced from Ref. \cite{sharma2026gravity}, (b) rectangular with sides $a$ and $2a$, and (c) equilateral triangular with side length $a$. A schematic of the geometry is also shown in (c). Marker shapes indicate the aperture shape used. In all cases, the flux is observed to approach an asymptotic value independent of particle size for $a / d \gg 1$. To gain some insight into the internal dynamics of the discharge, we next present complimentary simulations. \smallskip

%%%%%%%%%%%%%%%%%%%%%%%%%%%%%%%%%%%%%%%%%%%%%%%%%%%%%%%%%%%%%%%%%%%%%%%%
%%%%%%%%%%%%%%%%%%%%%%%%  SECTION III: Simulations  %%%%%%%%%%%%%%%%%%%%
%%%%%%%%%%%%%%%%%%%%%%%%%%%%%%%%%%%%%%%%%%%%%%%%%%%%%%%%%%%%%%%%%%%%%%%%

\section*{Simulations}

\begin{figure*}
\centering
\includegraphics[width =0.85\textwidth]{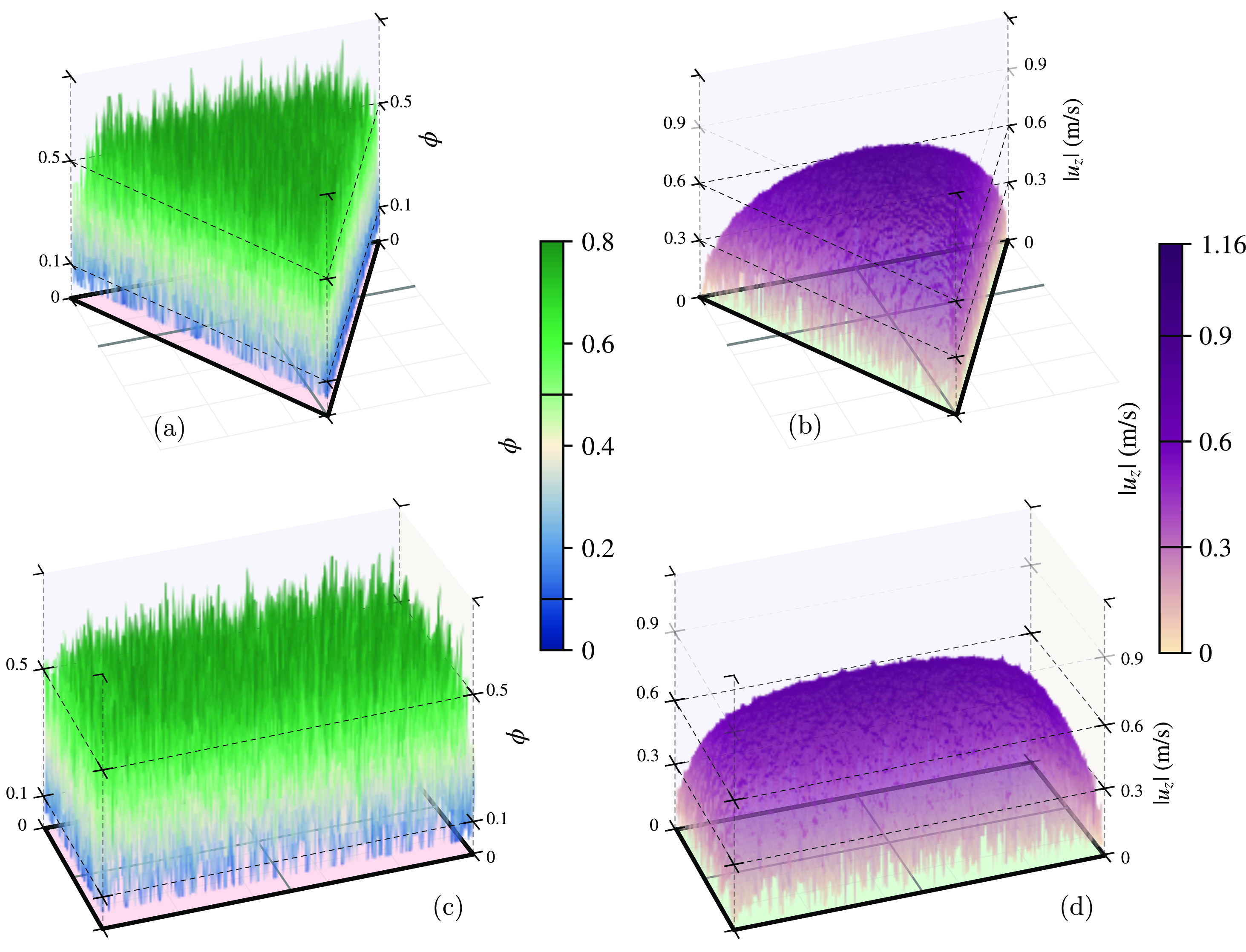}
\caption{Time-averaged fields at the aperture [plane at $z=0$], obtained from DEM simulations with $d=1 \, \text{mm}$. The upper row [(a)–(b)] shows fields for an equilateral triangular aperture of side $a=70\, \text{mm}$, and the lower row [(c)–(d)] shows a rectangular aperture of sides $a=30 \, \text{mm}$ and $2a$ [$D_h/d \simeq 40$ for both]. Panels (a),(c) show the packing fraction $\phi (x,y)$, while panels (b),(d) show the gravity-driven downward velocity, $u_z(x,y)$. Packing is approximately plug-like ($\phi_{\rm ff} \simeq 0.55$) away from a particle-scale depleted region adjacent to the boundary, whereas the velocity forms a smooth dome with a maximum near the aperture centroid. Surface height and color both represent the corresponding field magnitude averaged over the steady discharge duration.}
\label{fig:Fig3_Sim_oneExample}
\end{figure*}

We performed Discrete Element Method (DEM) simulations using the LIGGGHTS software package \cite{liggghts}. In these simulations, a flat-bottomed cylinder of diameter $20 \, \mathrm{cm}$ is filled with $7 \, \mathrm{kg}$ of spherical particles of a single diameter $d = 1$, $2$, or $4 \, \mathrm{mm}$ and density $\rho_\mathrm{g}=2500 \, \mathrm{kg}\cdot\mathrm{m}^{-3}$. We generate a single initial packing for each particle diameter. As in the experiments, the particles are then drained through a central aperture located at the bottom of the cylinder. For triangular apertures [made of three equal sides of length $a$], $a$ ranges from $20$--$80$ mm. For rectangular apertures [of sides $a$ and $2a$], $a$ ranges from $20$--$60$ mm. A complete list of the apertures and numerical parameters used in the simulations is provided in the S.M. (S.II.). \smallskip

During draining, Lagrangian data for particle positions and velocities are recorded at $10\,\mathrm{Hz}$. From these data, we compute the time-averaged Eulerian volume fraction and vertical velocity fields, $\phi(\mathbf{x})$ and $u_z(\mathbf{x})$ following the procedure detailed in the Supplemental Material. The average mass flux is computed as
\begin{equation}
    Q = \rho_\mathrm{g} A \langle u_z \phi\rangle_A,
\end{equation}
where, for an arbitrary field $f(\mathbf{x})$, $\langle\,\cdot\,\rangle_A$ denotes the plane-average
\begin{equation}
    \langle f \rangle_A = \frac{1}{A}\iint_{\mathcal{A}} f(x, y, 0)\,\mathrm{d}A,
    \label{eq:average_definition}
\end{equation}
taken over the aperture cross section $\mathcal{A}$, which lies in the plane $z=0$ and has area $|\mathcal{A}|=A$. We compute the average packing and vertical velocity over the aperture, $\langle\phi\rangle_A$ and $\langle u_z\rangle_A$, analogously. The simulation results for the steady flux $Q$ are shown in Fig.~S1 of the Supplemental Material. \smallskip

The highly symmetrical structure in the circular case allows for averaging over the angular coordinate, resulting in smooth profiles for packing and velocity averaged over space and time. For less ideal apertures, even averaging over the steady duration does not necessarily result in smoothed profiles. Further, to resolve sharp corners more accurately, we refine the mesh over which these measurements are taken. Consequently, the packing profiles remain grainy. In Fig.~\ref{fig:Fig3_Sim_oneExample}, we show example profiles of packing $\phi(\mathbf{x})$ [(a),(c)] and velocity $u_z(\mathbf{x})$ [(b),(d)] for each aperture shape. In Fig.~\ref{fig:Fig3_Sim_oneExample} (a)-(b), we report simulation profiles of an equilateral triangular aperture, with sides $a = 70 \, \mathrm{mm}$ and $d = 1 \, \mathrm{mm}$. An example rectangular aperture is shown in Fig.~\ref{fig:Fig3_Sim_oneExample} (c)-(d) with sides $a$ and $2a$, where $a = 30 \, \mathrm{mm}$ and $d = 1 \, \mathrm{mm}$. Despite the grainy texture of the packing profiles, it is clear that the two distributions exhibit distinct spatial structures. The packing profiles show some boundary-induced dilation, as the packing rapidly decreases at the boundaries [the diverging colormap is centered at $\phi=0.4$ as white]. Far from the edges, despite the grainy texture reflecting the discrete nature of discharge, the packing recovers a free-fall value close to $\phi_{\rm ff} \approx 0.55$. The velocity profiles, meanwhile, show a dome-shaped structure, with peak velocity near the aperture centroid. Both these observations are in close qualitative agreement with profiles observed in simulations of circular apertures \cite{sharma2026gravity}.

\begin{figure}
\centering
\includegraphics[width =\columnwidth]{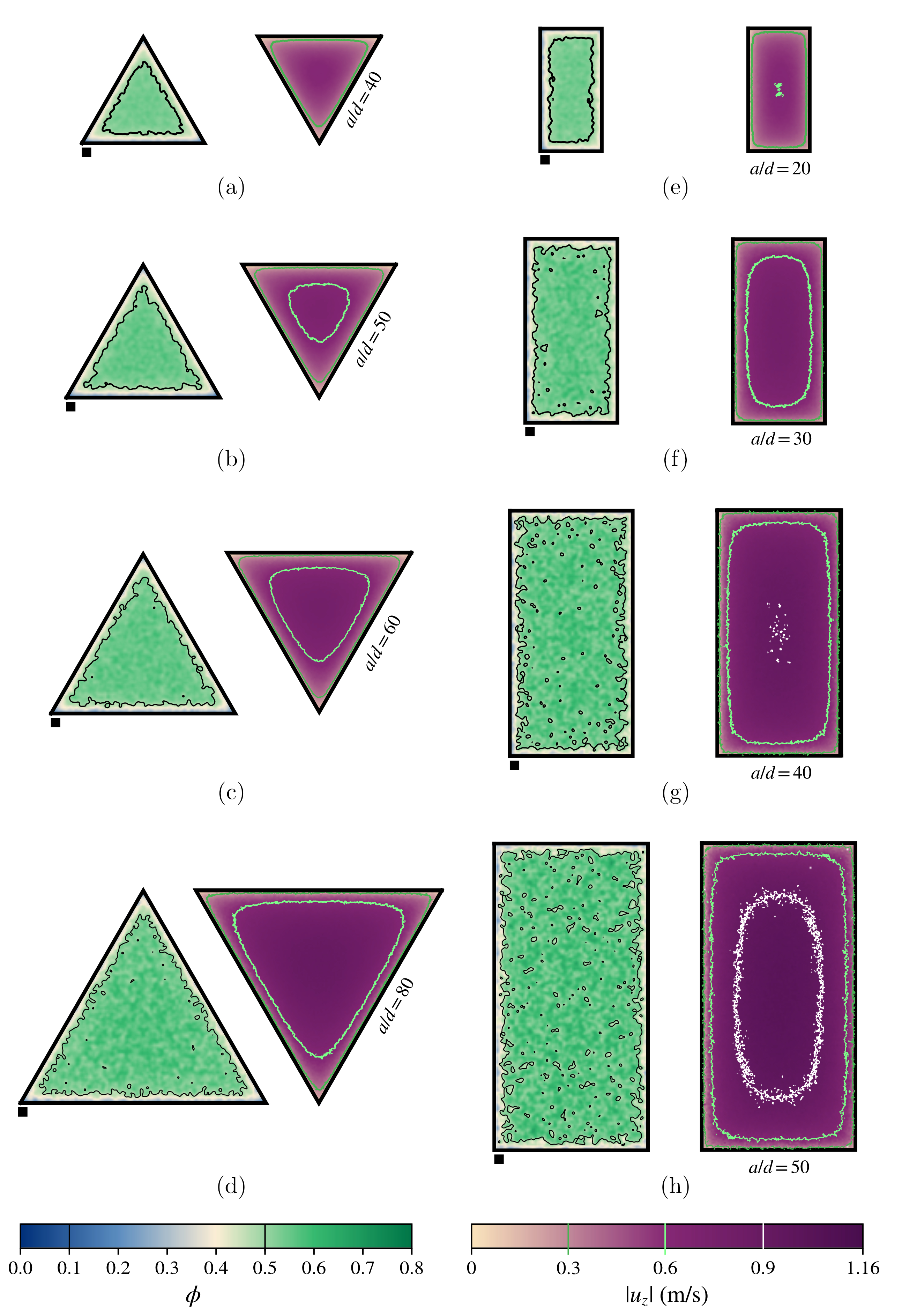}
\caption{The time-averaged data for $\phi(x,y)$ and $u_z(x,y)$ at the aperture for a range of relative aperture sizes $a/d$ for both the triangular (a)-(d) and rectangular (e)-(h) apertures during steady discharge, with $d = 1 \, \mathrm{mm}$. For each geometry, the fields are averaged using the rotational and reflection symmetries appropriate to the aperture, and a Gaussian blur of radius $d/2$ is applied, after which contour lines are drawn to emphasize the field structure. Across both shapes and sizes, a boundary layer of dilated packing persists at the wall with essentially unchanged absolute thickness as the aperture grows around it. Scale square shown of side $3d$. Meanwhile, the velocity profiles remain smooth over the aperture interior.}
\label{fig:Fig4_Sim_with_contours}
\end{figure}

In Fig.~\ref{fig:Fig4_Sim_with_contours}, we show packing and velocity profiles for a range of aperture sizes with $d = 1$ mm, with a colorbar indicating the local field magnitude. Here, we first perform spatial averaging based on the relevant symmetries of each shape, in addition to temporal averaging over the steady-flux region. To emphasize differences between the two fields, we overlay contour lines after Gaussian blurring with radius $d/2$, corresponding to the particle radius. The dilated region near the edges is largely independent of aperture shape and size. A small black square of side $3d$ is shown next to each of the packing profiles. We next discuss the packing and velocity structure. \smallskip

\subsection*{Packing at the aperture}

\begin{figure}
\centering
\includegraphics[width =\columnwidth]{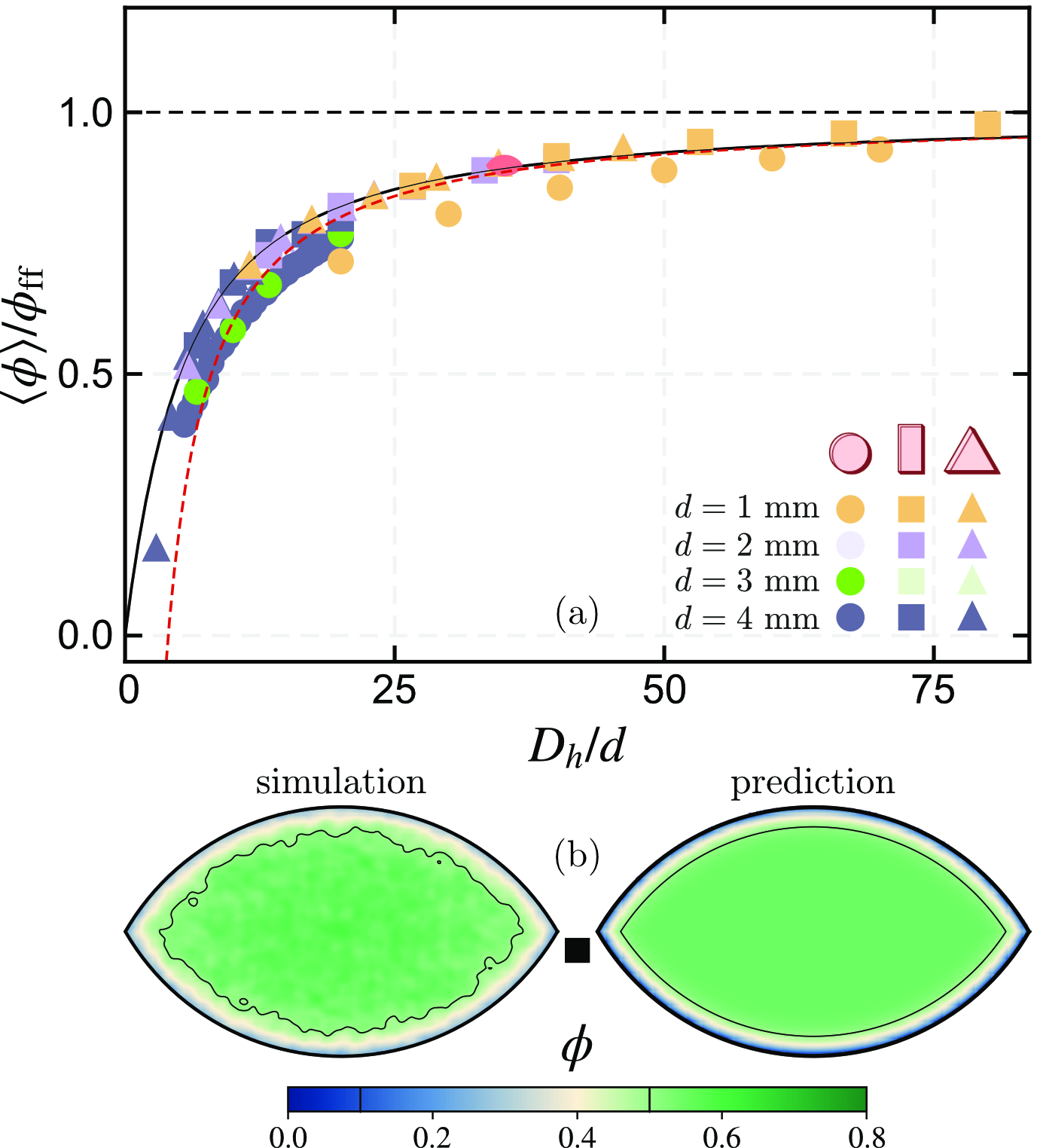}
\caption{(a) Aperture-averaged packing fraction for all aperture shapes, sizes, and grain sizes tested, as a function of relative aperture size from simulations. Marker shape denotes aperture geometry, and color denotes particle diameter. The black curve shows the prediction from Eq.~\eqref{eq:general_phi} for circular and equilateral triangular apertures, which coincide at fixed $D_h/d$ for $\ell=d$. The exact result for a 2:1 rectangle differs by a small higher-order correction. The red dashed curve shows the common leading large-aperture approximation (Eq.~\eqref{eq:phi_Dh}). (b) A lens-shaped aperture formed by the intersection of two circles of radius $R=30d$, with their centers separated by $R$, is simulated and compared with the packing profile predicted by Eq.~\eqref{eq:phi_s}, using the same $\ell=d$. The black square has side $3d$. Simulation (left) and prediction (right) agree closely, even without additional fitting.}
\label{fig:Fig5_Sim_Phi_mean}
\end{figure}

Across all aperture shapes tested, the time-averaged packing profiles $\phi(\textbf{x})$ share a common qualitative structure: a flat (albeit grainy) interior where $\phi (\textbf{x}) \simeq \phi_{\rm ff} \simeq 0.55$, bounded by a thin, particle-scale layer of dilation at the wall [Fig.~\ref{fig:Fig4_Sim_with_contours}]. In the limit that the aperture is much larger than the particles, the interior dominates. Therefore, for large apertures, we expect the mean packing over the entire aperture to approach the free-fall value: $\langle \phi\rangle_A \to \phi_{\rm ff}$. This mirrors the circular-aperture case, where the mean packing relaxes to the asymptotic free-fall value $\phi_{\rm ff} \simeq 0.55$ for spheres, corresponding to the loss of rigid contact structure at the random-loose-packing threshold \cite{onoda1990random, prado2013incompressible, sharma2026gravity}. This is a property of the granular packing and is expected to be unchanged for different aperture shapes.  The packing structure displays stark modification near the aperture edges. While this effect diminishes in relevance asymptotically for $a/d \gg 1$, the boundary-induced dilation contributes to flux suppression for finite apertures, as in the circular case. This is evident in the measurements of $\langle \phi \rangle_A / \phi_{\rm ff}$ across a range of relative sizes and shapes [Fig.~\ref{fig:Fig5_Sim_Phi_mean}]. \smallskip

Boundary-induced changes in packing over distances comparable to the particle size have also been reported for static confined packings, for instance in Ref.~\cite{desmond2009packing}. Whether a system is static or flowing, boundary exclusion is inescapable. The closest a spherical particle's center can get to a flat rigid boundary is its own radius, $d/2$. Further, if we consider an infinitesimal volume near such a boundary, $\phi \to 0$, as particles contact the boundary at individual points. Given $\phi \to 0$ at the boundary for spheres, and $\phi \to \phi_{\rm ff}$ far from the edge, in Ref.~\cite{sharma2026gravity} the recovery toward $\phi_{\rm ff}$ was modeled phenomenologically as a first-order spatial relaxation, \begin{equation}
    \frac{\dd\phi}{\dd s} =\frac{\phi_{\rm ff}-\phi(s)}{\ell},
    \label{eq:phi_relax}
\end{equation} where $\ell$ is a wall-normal recovery length, and $s$ is the distance from the nearest wall. Across relative aperture and particle sizes, $\ell \simeq 1d$ in Ref.~\cite{sharma2026gravity}. In words, rather than an empty annulus as suggested in earlier works like Ref.~\cite{brown1970principles}, the packing dilation is smooth provided adequate steady discharge can be averaged over. From the data in Ref.~\cite{sharma2026gravity}, the particle scale sets a characteristic length-scale over which the packing relaxes exponentially, rather than a literal empty annulus. This local relaxation is expected to be independent of aperture shape, provided the local radius of curvature much larger than $d$, away from corners. \smallskip

This suggests that the local packing is governed primarily by the shortest distance to the aperture boundary. Let $s (\textbf{x}) = \mathrm{dist} (\textbf{x}, \partial\mathcal{A})$ denote this shortest distance from a point $\textbf{x}$ in the aperture $\mathcal{A}$ to its boundary $\partial\mathcal{A}$. The resulting packing profile is found by integrating Eq.~\eqref{eq:phi_relax}
\begin{equation}
    \phi(s) = \phi_{\rm ff} \left[1 - {\rm exp} \left({-\frac{s}{\ell}}\right)\right].
    \label{eq:phi_s}
\end{equation} Rather than presenting a shape-dependent relaxation length, we consider $\ell = 1d$, as the only microscopic length-scale available to the system. This equation [Eq.~\eqref{eq:phi_s}] is local and geometric. Once the outline of an aperture is specified, the distance field $s(\textbf{x})$ determines the predicted packing field throughout the opening. It accounts only for the nearest boundary and therefore neglects additional influence from more distant walls or interactions between neighboring boundary layers near corners. Nevertheless, the measured triangular and rectangular fields in Fig.~\ref{fig:Fig4_Sim_with_contours} show the same wall-normal recovery, and we provide predictions using this local profile alongside these simulation measurements in the S.M. Fig.~S2. \smallskip 

We compare model predictions with simulation measurements by considering the average packing over the entire aperture. Using the aperture average defined in Eq.~\eqref{eq:average_definition}, the profile in Eq.~\eqref{eq:phi_s} gives \begin{equation}
    \langle\phi\rangle_A =
    \frac{1}{A}\int_{\mathcal{A}}\phi\, [s(\mathbf{x})]\,\mathrm{d}A.
    \label{eq:phi_avg}
\end{equation} To evaluate this average for an arbitrary shape, we define an inward-eroded region $A_s = \{\textbf{x} \in \mathcal{A}: \mathrm{dist} (\textbf{x}, \partial\mathcal{A}) \geq s\}$. This allows us to define $A(s) = \abs{A_s}$ as the area remaining after an inward displacement of $s$ from all boundaries. Thus, sitting at the boundary, no region is eroded, and $A(s=0)=A$. We define $s_{\rm max}$ as the greatest distance from any point in the aperture to its boundary. Thus, after having eroded $s = s_{\rm max}$ from each boundary, $A(s_{\rm max}) = 0$, \textit{i.e.}, no area remains. Putting Eqs.~\eqref{eq:phi_s} and \eqref{eq:phi_avg} together, the mean packing is \begin{equation}
    \frac{\langle\phi\rangle_A}{\phi_{\mathrm{ff}}}=1-\frac{1}{A}\int_0^{s_{\mathrm{max}}} \left(-\frac{\dd A(s)}{\dd s}\right) \left[\, \exp \Big({-\frac{s}{\ell}}\Big)\right] \dd s.
    \label{eq:general_phi}
\end{equation} The exponential term describes how the boundary influence relaxes into the interior, while $-\dd A /\dd s$ measures how much aperture area lies at a given distance from the boundary. At this stage, we assume no effective aperture size. \smallskip

For the geometries considered here, $A(s)$ can be obtained directly. For example, inward offsets of a circle or regular polygon remain geometrically similar to the original opening, giving
\begin{equation}
    A(s)=A\left(1-\frac{s}{s_{\mathrm{max}}}\right)^2.
\end{equation}
This, however, does not work for rectangles, as moving a fixed distance inwards modifies the aspect ratio of the rectangle; instead, it can be described as $A_{\square} (s) = (a -2s)(2a-2s)$ for $0 \leq s \leq a/2$. We evaluate Eq.~\eqref{eq:general_phi} using the appropriate $A(s)$ for each geometry and the resulting predictions are compared with the simulation measurements in Fig.~\ref{fig:Fig5_Sim_Phi_mean}(a) as a black line. When compared against the hydraulic diameter normalized by $\ell = 1d$, the circle and equilateral triangle give identical predictions, whereas the $2{:}1$ rectangle exhibits a small difference at higher order, while the three curves are indistinguishable in Fig.~\ref{fig:Fig5_Sim_Phi_mean}(a). Across aperture shape, aperture size, and grain size, the model captures the effect of confinement-induced dilation. \smallskip

The geometric interpretation of Eq. \eqref{eq:general_phi} becomes particularly transparent in the large-aperture limit, where the three profiles share the same leading order correction. Because the exponential confines the packing deficit to $s = \mathcal{O} (\ell)$, only the behavior of $A(s)$ near $s=0$ contributes when $s_{\rm max} \gg \ell$, \textit{i.e.}, the aperture is large compared to the particle. For a convex aperture with a piecewise smooth boundary,
\begin{equation}
    A(s)=A-Ps+\mathcal{O}(s^2),
    \label{eq_As_higherOrder}
\end{equation} where $P$ is the perimeter. The expansion is taken at fixed aperture shape and aspect ratio as its overall size increases. It does not need to be uniform for arbitrarily sharp corners or increasingly slender openings. Hence $-(\dd A/\dd s) = P + \mathcal{O}(s)$, and thus from Eq.~\eqref{eq:general_phi},
\begin{equation}
    \frac{\langle\phi\rangle_A}{\phi_{\rm ff}} = 1-\frac{P\ell}{A} + \mathcal{O}\Big(\left[\frac{\ell}{L}\right]^2\Big).
    \label{eq:meanPacking_largeDh/d}
\end{equation} The leading packing deficit therefore depends only on the boundary layer area $P\ell$ relative to the total aperture area $A$. Equivalently, this term can be interpreted as a ratio that measures the relevance of boundary-controlled particles to the aperture as a whole. Written as $(P/\ell) / (A/\ell^2)$, the numerator $P/\ell$ estimates the number of particle-scale regions required to trace the boundary, while $A/\ell^2$ estimates the number required to fill the aperture. Edge effects enter at first order in $\ell/L$, where $L$ is a physical length-scale associated with the aperture, while corner-specific corrections enter through the higher-order terms in Eq.~\eqref{eq_As_higherOrder}. A hydraulic diameter like measure emerges as the leading confinement length from the local relaxation of packing, described over an aperture. It measures the aperture interior available per unit boundary and thus determines what fraction of the opening is affected by the particle-scale depleted layer. \smallskip

Introducing the hydraulic diameter, $D_h \equiv 4A/P$, Eq.~\eqref{eq:meanPacking_largeDh/d} becomes
\begin{equation}
    \frac{\langle\phi\rangle_A}{\phi_{\mathrm{ff}}} \simeq 1-\frac{4\ell}{D_{h}}.
    \label{eq:phi_Dh}
\end{equation} This large-aperture prediction is shown by the dashed red curve in Fig.~\ref{fig:Fig5_Sim_Phi_mean}(a) with $\ell = 1d$. It approaches the complete geometric calculation from Eq.~\eqref{eq:general_phi} as $D_h/d$ increases, while departures at stronger confinement retain information about corners and the full form of $A(s)$. \smallskip

Finally, we test whether the local construction is predictive beyond the three aperture families used to establish the result. Fig.~\ref{fig:Fig5_Sim_Phi_mean} (b) considers a regular lens-shaped aperture formed by the intersection of two equal circles, with the center of each circle lying on the circumference of the other. Using only the aperture outline and the same relaxation length $\ell = 1d$, Eq.~\eqref{eq:phi_s} predicts the two-dimensional packing field. The measured and predicted fields agree closely, including the depleted layer that follows the curved boundary and the recovery toward a nearly uniform interior, without any additional fitting. The average packing is also plotted in Fig.~\ref{fig:Fig5_Sim_Phi_mean}(a) with a red marker. The lens-shaped test is chosen since it is neither one of the polygonal geometries used in the primary study nor a small perturbation of a circle. Its agreement supports the view that the packing deficit is a local response to the nearest boundary over a few particle diameters, while the hydraulic diameter arises after integrating that local response over the full aperture. \smallskip

\begin{figure}
\centering
\includegraphics[width =\columnwidth]{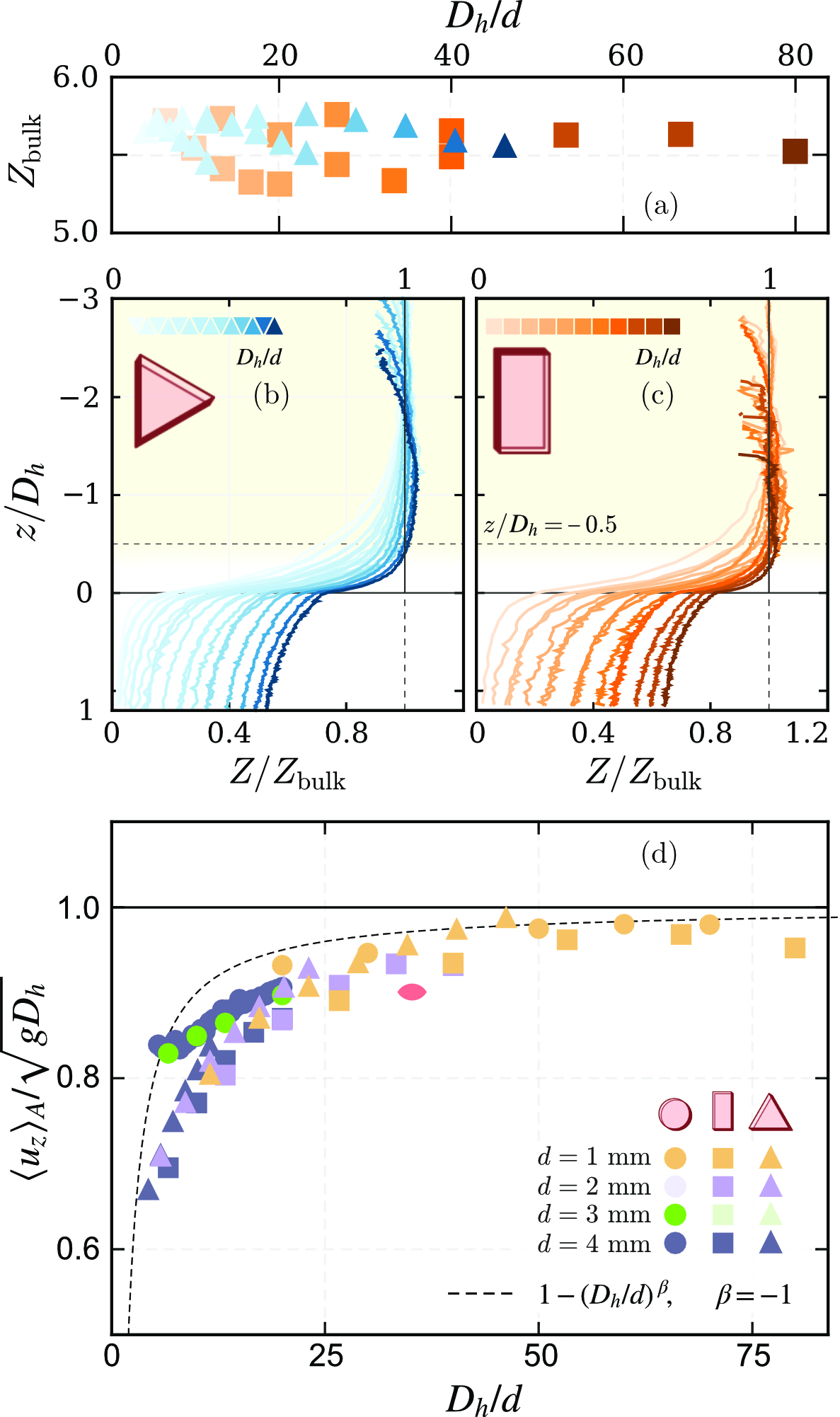}
\caption{(a) Bulk coordination number $Z_{\rm bulk}$ defined as the mean grain coordination within a region in the container,$-2<z/D_h<-1$, above the aperture. Mean coordination number $Z$, normalized by $Z_{\rm bulk}$ for (b) triangular and (c) rectangular apertures. The coordination number decreases continuously toward and beyond the outlet. For $D_h \gg d$, the profiles depart from their bulk values about $D_h/2$ above the aperture, as indicated by the horizontal dashed line. (d) Aperture-averaged velocity normalized by the resulting free-fall scale, $\langle u_z\rangle_A/\sqrt{gD_h}$, as a function of $D_h/d$. The normalized velocity from all simulations approaches unity for large $D_h/d$ for all shapes. The dashed curve, $1-(d/D_h)$, is shown only as a guide to the relaxation toward the asymptote.}
\label{fig:Fig6_contactsVelocity}
\end{figure}

\subsection*{Velocity in free-fall}

The detailed velocity field at the aperture depends on the geometry of the opening and, for finite $D_h/d$, may also retain the influence of particle-scale contact mechanics. The velocity profiles [Fig.~\ref{fig:Fig3_Sim_oneExample} (b), (d) and Fig.~\ref{fig:Fig4_Sim_with_contours}], similar to the packing going to $0$ at the edges, much more gradually than their packing counterpart. These ingredients determine how the flow approaches its asymptotic state. In the limit $D_h/d\gg1$, however, following the circular case, we expect the leading velocity scale to become independent of the grain size and be set primarily by gravity and some characteristic length associated with the aperture. Rather than provide a local model for the velocity across aperture shapes, we focus here on the aperture-averaged downward velocity, $\langle u_z \rangle_A$, rather than seeking a universal local velocity profile for arbitrary shapes. \smallskip

For circular apertures, $D/2$ provides the classical free-fall-arch length scale. Ref.~\cite{rubio2015disentangling}, however, showed that contact pressure decreases continuously rather than vanishing across a sharply defined arch. Nevertheless, this characteristic length scale provides a kinematic reference for the mean discharge velocity at the aperture in the circular setting. To identify the relevant acceleration length, we examine the mean coordination number per particle, $Z$, from simulations inside the vertical column and averaging over the steady discharge window $\Delta T$. Using the particle positions, neighbors are counted in $Z$ if the surface gap $< 0.02d$ apart in that frame. The column is vertically binned with height $\Delta z=0.5d$. $z = 0$ is defined as the aperture plane, with $z<0$ inside the container and $z>0$ below the aperture. We define $Z_{\mathrm{bulk}}$ as the mean coordination number in a region above the aperture, $-2<z/D_h<-1$, during steady discharge. Despite some small dependence on particle and aperture size, across simulations $Z_{\rm bulk} \simeq 5.5$ [Fig.~\ref{fig:Fig6_contactsVelocity}(a)]. In Figs.~\ref{fig:Fig6_contactsVelocity}(b) and (c), $Z(z)$ is normalized by its corresponding bulk value for triangular and rectangular apertures, respectively. Rather than vanishing abruptly, the coordination number decreases gradually as the grains approach the outlet. The transition from a contact-supported column to the emerging jet is distributed over a finite vertical region. For both aperture shapes, the coordination number $Z$ begins to decrease approximately a distance $D_h/2$ above the outlet, at $z\simeq-D_h/2$, particularly for $D_h\gg d$. \smallskip

\begin{figure*}
\centering
\includegraphics[width =\textwidth]{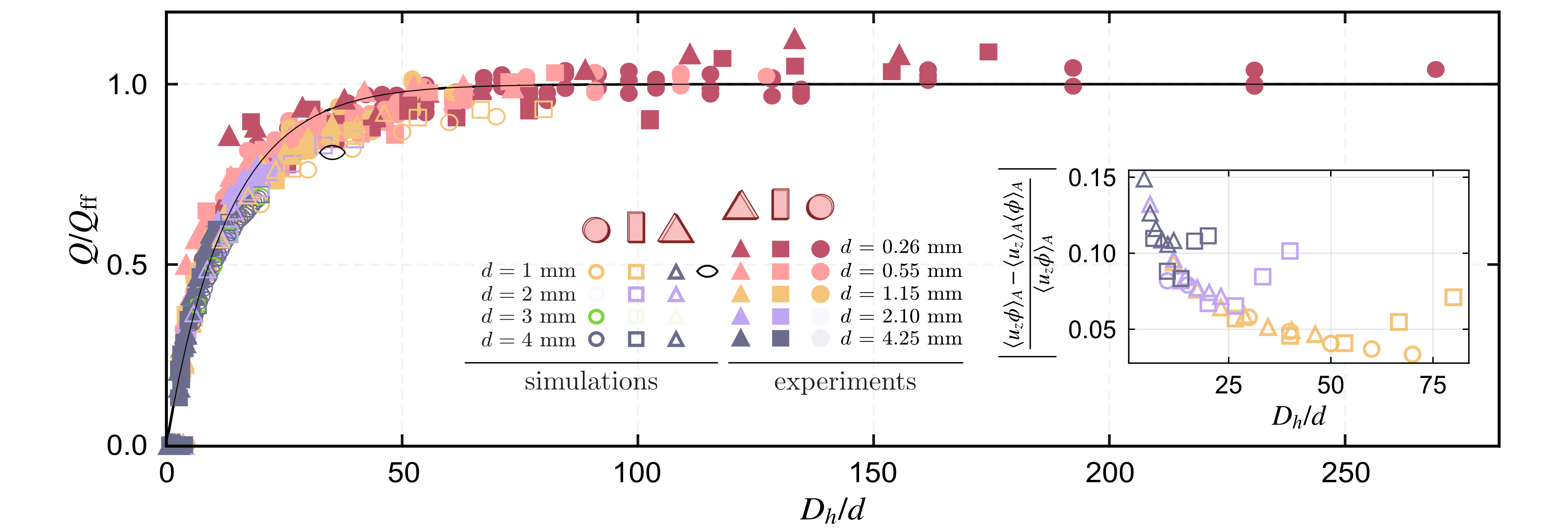}
\caption{Dimensionless flux $F = Q/Q_{\rm ff}$ as a function of relative aperture size, across all aperture shapes (marker shape), particle sizes (marker color), experiments (filled markers), and simulations (open markers). The results for a circular aperture from Ref.~\cite{sharma2026gravity} are included for comparison. Data from the triangular and rectangular apertures collapse onto the circular-aperture trend when using the hydraulic diameter $D_h$ as the relevant length scale. The solid line shows the one-parameter form of Eq.~\eqref{eq:oneParameter_F}, with $nd = 13d$. Simulations relax toward the asymptotic value over more grain diameters than experiments, though both are well described by the same functional form; individual best fits across grain size and aperture shape fall in the range $n = 10$-$15$, consistent with Ref.~\cite{sharma2026gravity}. Simulation results with a lens-shaped aperture are shown with a black marker. The inset shows the error by assuming $\langle u_z \phi\rangle_A=\langle u_z\rangle_A\langle \phi\rangle_A$. The error appears to increase as $D_h/d$ increases for rectangular apertures, and this is likely caused by the limited number of steady-state frames for large apertures. Aside from these cases, the error decreases with increasing $D_h/d$, consistent with Ref.~\cite{sharma2026gravity}.}
\label{fig:Fig7_Collapse}
\end{figure*}

This suggests $L \simeq D_h/2$ may also serve as the characteristic distance over which the contact-supported flow transitions to free fall. A kinematic estimate over this distance gives $\langle u_z \rangle_{A, \, {\rm ff}} \simeq \sqrt{2gL} = \sqrt{gD_h}$. This estimate assumes negligible entry velocity and negligible mean axial contact force over $L$. Here, free fall denotes an asymptotic kinematic reference rather than a sudden transition across a physical arch. We test this proposal for the relevant transport length directly in Fig.~\ref{fig:Fig6_contactsVelocity}(d), where the aperture-averaged velocity is normalized by $\sqrt{gD_h}$. Across circular, triangular, and rectangular apertures, and over the simulated range of particle sizes, $\langle u_z \rangle_A / \sqrt{gD_h} \to 1$ as $D_h/d$ is increased. The dashed curve in Fig.~\ref{fig:Fig6_contactsVelocity}(d) is included only as a guide to the finite-confinement relaxation and is not intended as a universal correction. 
In the circular case, even the smallest $D/d$ apertures with steady flows yield a mean velocity $\approx 0.8 \sqrt{gD}$; however, finite flows with alternative aperture shapes show smaller mean velocities [$\approx 0.7\sqrt{gD_h}$]. Nonetheless, for large $D_h/d$, the variation diminishes progressively. In other words, geometric and mechanical details may govern the relaxation toward this free-fall state at finite $D_h/d$. Still, their influence on the leading velocity scale should diminish as the aperture becomes much larger than the particles. Overall, the hydraulic diameter provides a robust transport length across tested shapes, and is verified for the example case of the lens-shaped aperture (marked in red) with $D_h/d \approx 35$, for which $\langle u_z \rangle_A \simeq 0.9\sqrt{gD_h}$. \smallskip

Unlike the packing field, whose local variation can be expressed using the distance from the nearest boundary, the pointwise velocity profile derived previously for circular apertures relied on the aperture radius \cite{sharma2026gravity} and has no unique analog for a general opening. We therefore restrict the present result to the aperture-averaged velocity. The hydraulic diameter nevertheless plays complementary roles in the two fields: it measures the relative importance of the particle-scale boundary layer for packing, while empirically providing a characteristic release length consistent with the measured mean-velocity scale.

\section*{Discussion}

The measured discharge rates in Fig.~\ref{fig:Fig1_appertureFlux_raw} provide an integrated measure of the outlet flow:
\begin{equation}
    Q = A \, \rho_g \, \langle u_z \, \phi \rangle_A.
\end{equation} The density of the grains ($\rho_g$) trivially survives modification in aperture shape. Likewise, the aperture area ($A$) is well defined for any 2D-shaped aperture. Both experimental and numerical values of $Q$ do not automatically resolve the underlying packing and velocity distributions, rather just a scalar mean. For the combined distribution, averaging spatially over $A$ [denoted $\langle \cdot \rangle_A$] and temporally over the steady duration $t$ [denoted by overline], the order of averaging makes no difference, $\overline{\langle \phi u_z\rangle_{A}} = \langle \overline{\phi u_z}\rangle_{A}$. \smallskip

In attempting to separate an independent mean velocity scale, we must decide whether spatial or temporal averaging is considered first, and where packing normalization is conducted. This leads to two different definitions of covariance. First, (i) a temporal covariance. We consider the time-varying, aperture-averaged packing $p(t) \equiv \langle \phi(t) \rangle_A$. Consequently, we define an occupancy-weighted aperture velocity $v (t) \equiv \langle \phi u_z (t)\rangle_A \, / p(t)$. The temporal covariance is defined as \begin{equation}
    C_t \equiv \overline{p v} \,\, - \,\, \overline{p} \,  \overline{v}.
\end{equation} In Ref. \cite{sharma2026gravity}, this was the covariance measured despite the unspecified notation. It was found to be a fraction of a percent for circular apertures, in particular for $D \geq 20d$.  A similarly negligible contribution is observed for the other aperture shapes here when $D_h \gg d$.  \smallskip  

Additionally, here we consider (ii) the spatial covariance between the two distributions. We define a spatially-varying, time-averaged packing $\Phi(\textbf{x}) \equiv \overline{\phi(\textbf{x})}$. Thus, the velocity $V(\textbf{x}) = \overline{\phi u_z (\textbf{x})} / {\Phi(\textbf{x})}$. This is the velocity discussed in the velocity section earlier, and details are provided in S.M. The spatial covariance, \begin{equation}
    C_A \equiv \langle \Phi V\rangle_A   \,\, - \,\, \langle \Phi \rangle_A\langle V \rangle_A.
\end{equation} is shown in the inset of Fig.~\ref{fig:Fig7_Collapse} as a function of $D_h/d$. Data from across our simulations is shown as the percentage correction, $C_A / (Q/\rho_g A)$. Some data from Ref.~\cite{sharma2026gravity} for circular apertures is re-analyzed and also shown here. For smaller $D_h/d$, the relative correction is considerably larger than the temporal counterpart, with corrections as large as $\sim 15\%$ for severely constrained flows. While these trends consistently show that larger apertures contribute a smaller percentage correction, for the large rectangles this trend appears to reverse. This is a likely due fast discharges with a limited overall mass where we can have only $10$ and fewer temporal snapshots in the steady regime. From the two different covariances, the aperture-integrated flow is steady enough for the temporal correction to be negligible, while the persistent spatial structure [lower packing and velocity near the edges] produces a modest spatial correction. 

Regardless of definition choice, for relatively large apertures, the covariance corrections are small and the mean product approximately factorizes, $\langle \phi u_z \rangle_A \approx \langle \phi \rangle_A\langle u_z\rangle_A$. The free fall state can consequently be interpreted by the asymptotic cases of the packing and velocity distributions individually as $D_h /d \to \infty$. Thus, the free-fall flux is \begin{equation}
    Q_{\rm ff} = A \, \rho_g \, \phi_{\rm ff} \,  \sqrt{gD_h}
    \label{eq:flux_FF_genApp}
\end{equation} where $\phi_{\rm ff} \simeq 0.55$ for nearly monodisperse spheres independent of the aperture shape. This is shown in a solid line for each aperture in Fig.~\ref{fig:Fig1_appertureFlux_raw}. Changes in aperture shape mainly affect the velocity scale, where the primary transport length now corresponds to the hydraulic diameter. If the aperture were circular, this returns the free-fall reference flux \eqref{eq:Q_ff_MFI} presented in Ref.~\cite{sharma2026gravity}. \smallskip

Deviations from this free-fall state are driven by boundary confinement for finite $D_h/d$. A measure of the overall deviation is provided through the fraction $F \equiv Q / Q_{\rm ff}$, compared across experiments (filled) and simulations (empty) for aperture sizes and shapes, as well as different grain sizes in Fig.~\ref{fig:Fig7_Collapse}. Marker shapes indicate that the modification of aperture shape is well accounted for by the relative size using the hydraulic diameter $D_h/d$, as expected from the packing discussion, and the experiments collapse onto the circular-apertured cases very well. The simulations for the different shapes, similar to the circular ones, relax to the asymptotic value more slowly than the experiments. The simulation results are, however, themselves very similar regardless of aperture shape. A one-parameter expression is presented to capture the confinement of flux, combining both packing and velocity
\begin{equation}
    F = \frac{\langle \phi \, u_z\rangle_A}{\phi_{\rm ff} \, \sqrt{gD_h}} \simeq 1 - \exp\!\left(-\frac{D_h}{nd}\right).
    \label{eq:oneParameter_F}
\end{equation} This empirical exponential relaxation is chosen to capture the  rapid variation with $D_h/d$ where confinement dominates, all the way to an asymptotic approach to the free-fall state. As the insets showed, for the highly confined cases, separating the packing and velocity distributions requires some modeling for their covariance. Overall a single expression is useful in capturing the cumulative effects of finiteness. \smallskip

Earlier work also introduced an exponential correction to Beverloo's empirical relation and suggested a connection with the grain density near the outlet \cite{mankoc2007flow, beverloo1961flow}. A very comparable exponential has also been shown to capture the incompressible-compressible transition for granular jets when grains are poured through a funnel \cite{prado2013incompressible}. We plot equation (\ref{eq:oneParameter_F}) together with the data in Fig.~\ref{fig:Fig7_Collapse} in a solid black line using $n=13$ \cite{sharma2026gravity}. The best fit for the simulation data is with $n \simeq 15$. The data for each grain-size and aperture-shape combination for the experiments have individual best fits in the range $nd = 10d$ to $15d$, mirroring the results for circular apertures \cite{sharma2026gravity}. Altogether, the one-parameter expression for the flux with the exponential correction is robust across the shape and sizes of the apertures, and sizes of the grains tested:
\begin{equation}
    Q \simeq A \, \rho_g \sqrt{gD_h} \, \phi_{\rm ff} \, \Big( 1 - e^{-D_h/nd} \Big).
    \label{eq:eq_fluxNew}
\end{equation} The number of grain diameters required for relaxation to the asymptotic value is likely to depend on the distribution of grain sizes, and our simulations are made of strictly monodisperse spheres. In experiments, about $10\%$ variation in size around the mean is observed. The present comparison does not isolate this effect from differences in contact parameters, aperture thickness, or container geometry. Experiments on bidisperse mixtures show that the discharge rate does indeed depend on mixture composition \cite{benyamine2014discharge}.

\section*{Conclusion}

We examined steady granular discharge through apertures whose geometry extends beyond the commonly considered circular opening. Despite substantial differences in the local shape of the packing and velocity fields, the discharge through circular, rectangular, and triangular apertures is organized by a common geometric length: the hydraulic diameter, $D_h = 4A/P$. For a circular aperture, $D_h=D$, and the free-fall framework introduced in Ref.~\cite{sharma2026gravity} is recovered exactly. For non-circular apertures, $D_h$ accounts for the competing influence of the aperture interior, represented by its area $A$, and its confining boundary, represented by its perimeter $P$.\smallskip

The packing field provides a geometric origin for this length. The exclusion enforced by the aperture edge is local: the deficit near any boundary depends primarily on distance to the nearest wall. The aperture boundary imposes a depleted layer whose structure relaxes over a particle-scale distance $\ell \simeq 1d$. Averaging this local relaxation over the opening yields, for sufficiently large apertures, Eq.~\eqref{eq:meanPacking_largeDh/d}. $P\ell/A$ measures the fraction of the aperture influenced by its boundary, or equivalently the number of particle-scale regions required to resolve the boundary relative to those filling the aperture interior. The same local model predicts the packing field of an additional lens-shaped aperture without additional fitting, supporting its interpretation as a geometric boundary-layer description rather than a shape-specific construction.

Our simulations also find that the hydraulic diameter determines the velocity scale in the free-fall limit. The coordination number decreases continuously as the grains approach the opening. For both triangular and rectangular apertures, the contact structure begins to depart from its bulk state about a distance $D_h/2$ above the outlet, particularly when $D_h \gg d$. This identifies a characteristic release length $L \simeq D_h/2$, from which free-fall kinematics gives $\langle u_z \rangle_{A, \, {\rm ff}} \simeq \sqrt{gD_h}$. Particle-scale geometry and mechanics may determine how the velocity relaxes toward this state at finite $D_
h/d$, but gravity and aperture geometry determine the asymptotic velocity. \smallskip

Combining the asymptotic packing ($\phi_{\rm ff} \simeq 0.55$ for spheres) and velocity gives the generalized free-fall flux [Eq.~\eqref{eq:flux_FF_genApp}]. Across aperture shapes, grain sizes, experiments, and simulations, deviations from this limit collapse when expressed as a function of $D_h/d$ [Fig.~\ref{fig:Fig7_Collapse}]. The one-parameter form [Eq.~\eqref{eq:oneParameter_F}] with relaxation to the asymptote over $n \simeq 10$-$15$ particles retains the same scale found for circular apertures \cite{sharma2026gravity}. The resulting expression, Eq.~\eqref{eq:eq_fluxNew}, therefore describes steady discharge across the tested aperture geometries without introducing any additional shape-dependent fitting parameters. \smallskip 

Clogging in the present experiments is a separate stochastic outcome observed for small relative aperture and is not described by Eq.~\eqref{eq:eq_fluxNew}. The flux analysis presented in this study concerns steady flux and requires the granular discharge to last long enough. Because studying clogging in such systems would require many experiments to reach statistically relevant results, we do not infer a critical opening size or clogging probability from the present flux relation. Many additional complications, such as inter-particle adhesion, variations in particle shape, polydisperse particle assemblies, etc., are often present in real silo flows and cause deviations from the free-fall state; these remain to be analyzed in the present framework. Experiments with controlled cohesion provide a route to test this extension \cite{sharma2025models, gans2024discharge}. \smallskip 

Together with Ref.~\cite{sharma2026gravity}, these results provide a kinematic reference for steady discharge through the tested aperture geometries. In the limit that the aperture is much larger than the constituent particles, the integrated structure of discharge at the aperture has been described kinematically. A complete description of the dense jet beyond the aperture remains to be supplied, and this will be the subject of our next study.

\begin{acknowledgments}
The authors acknowledge support from the Gordon and Betty Moore Foundation (Grant DOI 10.37807/GBMF13831), the U.S. Army Research Office under Grant No. W911NF-23-2-0046, and NSF CBET 2533460.

\smallskip

The authors used ChatGPT and Codex (OpenAI; 5.6 and 6) for manuscript revision, literature searches, and checks of scientific consistency. The authors reviewed and corrected the AI-assisted material and take full responsibility for the manuscript.
\end{acknowledgments}

\section*{Author Contributions}
K.Y.Z.: Investigation, Methodology, Formal analysis, Writing---original draft.
R.S.S.: Conceptualization, Investigation (lead), Visualization, Formal analysis, Data curation, Writing---original draft, Writing---review \& editing.
A.L.: Software, Methodology, Writing---review \& editing.
E.M.: Funding acquisition, Writing---review \& editing.
A.S.: Conceptualization, Supervision, Project administration, Funding acquisition, Writing---review \& editing.

\section*{Data Availability}
The data supporting this article are available in [repository].

% Embedded REVTeX bibliography: no BibTeX run required.
%apsrev4-2.bst 2019-01-14 (MD) hand-edited version of apsrev4-1.bst
%Control: key (0)
%Control: author (72) initials jnrlst
%Control: editor formatted (1) identically to author
%Control: production of article title (-1) disabled
%Control: page (0) single
%Control: year (1) truncated
%Control: production of eprint (0) enabled
%

% Supplement appended as LaTeX source; citations share the bibliography above.
\clearpage
\onecolumngrid
\begin{center}
{\large\bfseries Supplementary Materials}\\[0.5em]
{\large Steady granular discharge across aperture shapes}
\end{center}
\twocolumngrid
\setcounter{section}{0}
\setcounter{subsection}{0}
\setcounter{equation}{0}
\setcounter{figure}{0}
\setcounter{table}{0}
\renewcommand{\thesection}{S.\Roman{section}}
\renewcommand{\thesubsection}{\thesection.\arabic{subsection}}
\renewcommand{\theequation}{s\arabic{equation}}
\renewcommand{\thefigure}{S\arabic{figure}}
\renewcommand{\thetable}{S\arabic{table}}
\providecommand{\RSS}[1]{\textcolor{purple}{#1}}
\providecommand{\KZ}[1]{\textcolor{teal}{#1}}
\providecommand{\AS}[1]{\begingroup\color{olive}#1\endgroup}

%%%%%%%%%%%%%%%%%%%%%% Details of Experiments %%%%%%%%%%%%%%%%%%%

\section{Details of Experiments}

The experiments were performed with spherical glass beads (Sigmund Lindner GmbH) of five mean diameters, $d=0.26$, $0.55$, $1.15$, $2.10$, and $4.25\,\mathrm{mm}$. Each bead population has a standard deviation in diameter of approximately $10\%$ of its mean diameter. The grain density ranges from $\rho_g=2500$ to $2560\,\mathrm{kg\,m^{-3}}$, and the static packing fraction inside the filled cylinder is $\phi_{\rm cyl}=0.60$--$0.61$. Table~\ref{tab:particle_properties} lists the mean diameter, standard deviation, density, and packing fraction for each bead population.

\begin{table}[h!]
\centering
\begin{tabular}{||c|c|c|c||}
\hline \hline
$d$ [mm] & $d_{\rm std. \, dev}$ [mm]& $\rho_g$ [kg/m$^3$] & $\phi_{\rm cyl}$ [adim] \\
\hline \hline
0.26 & 0.03 & 2520 & 0.61 \\
\hline
0.55 & 0.06 & 2500 & 0.60 \\
\hline
1.15 & 0.12 & 2540 & 0.60 \\
\hline
2.10 & 0.21 & 2560 & 0.61 \\
\hline
4.25 & 0.40 & 2510 & 0.60 \\
\hline \hline
\end{tabular}
\caption{Physical properties of the spherical glass particles used in the experiments. Here, $d_{\rm std.\,dev}$ is the standard deviation
of particle diameter, and $\phi_{\rm cyl}$ is the
packing fraction inside the filled cylinder.}
\label{tab:particle_properties}
\end{table}
In this study, we use various aperture sizes for two aperture shapes. We place (i) an equilateral triangle (denoted $\Delta$) with side $a$ and (ii) a rectangle (denoted $\square$) with dimensions $a$ and $2a$ at the base of the column before any experiment begins. Ref.~\cite {sharma2026gravity} tested several aperture materials and found no noticeable effects on the overall flux; thus, for this study, we made apertures only using a laser cutter on PMMA sheets with the thicknesses listed in Table~\ref{tab:aperture_details}.

\begin{table}
\centering
\begin{tabular}{||c|c|c|c||}
\hline \hline
\textbf{Shape} & \textbf{$a$ [mm]} & \textbf{$D_h$ [mm]}& \textbf{Thick. [mm]} \\
\hline \hline
$\Delta$ & 4.0 & 2.31 & 1.6 \\
$\Delta$ & 5.0 & 2.89 & 1.6 \\
$\Delta$ & 6.0 & 3.46 & 1.6 \\
$\Delta$ & 7.0 & 4.04 & 1.6 \\
$\Delta$ & 8.5 & 4.91 & 1.6 \\
$\Delta$ & 10.0 & 5.77 & 1.6 \\
$\Delta$ & 13.0 & 7.51 &  1.7 \\
$\Delta$ & 17.0 & 9.81 & 1.6 \\
$\Delta$ & 20.0 & 11.55 & 1.7 \\
$\Delta$ & 25.0 & 14.43 & 1.6 \\
$\Delta$ & 30.0 & 17.32 & 3.1 \\
$\Delta$ & 40.0 & 23.09 & 3.1 \\
$\Delta$ & 50.0 & 28.87 & 3.2 \\
$\Delta$ & 60.0 & 34.64 & 3.2 \\
$\Delta$ & 70.0 & 40.41 & 3.1 \\
\hline \hline
$\square$ & 2.5 & 3.33 & 1.6 \\
$\square$ & 3.5 & 4.67 & 1.6 \\
$\square$ & 5.0 & 6.67 & 1.6 \\
$\square$ & 6.0 & 8.00 & 1.6 \\
$\square$ & 7.0 & 9.33 & 1.7 \\
$\square$ & 8.5 & 11.33 & 1.6 \\
$\square$ & 10.0 & 13.33 & 1.6 \\
$\square$ & 12.0 & 16.00 & 1.6 \\
$\square$ & 15.0 & 20.00 & 1.7 \\
$\square$ & 17.0 & 22.67  & 3.1 \\
$\square$ & 20.0 & 26.67 & 1.6 \\
$\square$ & 23.0 & 30.67 & 3.1 \\
$\square$ & 26.0 & 34.67 & 3.2 \\
$\square$ & 30.0 & 40.00 & 3.1 \\
$\square$ & 34.0 & 45.33 & 3.2 \\
\hline \hline
\end{tabular}
\caption{Details of apertures used in the experiments.}
\label{tab:aperture_details}
\end{table}

%%%%%%%%%%%%%%%%%%%%%% Details of simulations %%%%%%%%%%%%%%%%%%%

\section{Details of Simulations}
Similarly, in the simulations, we use equilateral triangles with side length $a$ and rectangles with side lengths $a$ and $2a$ as the aperture. Table~\ref{tab:aperture_details_simulations} shows the values of $ a$, which cover a wide range of $D_h$.

During simulations, the Hertz-Mindlin model with tangential history and rolling friction is used for particle-particle and wall-particle interactions. The mechanical parameters are chosen to efficiently simulate glass beads of $d=\mathcal{O}(1\text{mm})$: Young's modulus, $E=1\, \text{MPa}$, Poisson ratio, $\nu=0.2$, coefficient of restitution, $e=0.5$,  sliding friction coefficient, $\mu_\mathrm{s}=0.5$, and coefficient of rolling friction, $\mu_\mathrm{R}=0.01$ \cite{dumont_micro_params_2020}.

\begin{figure}
\centering
\includegraphics[width =\columnwidth]{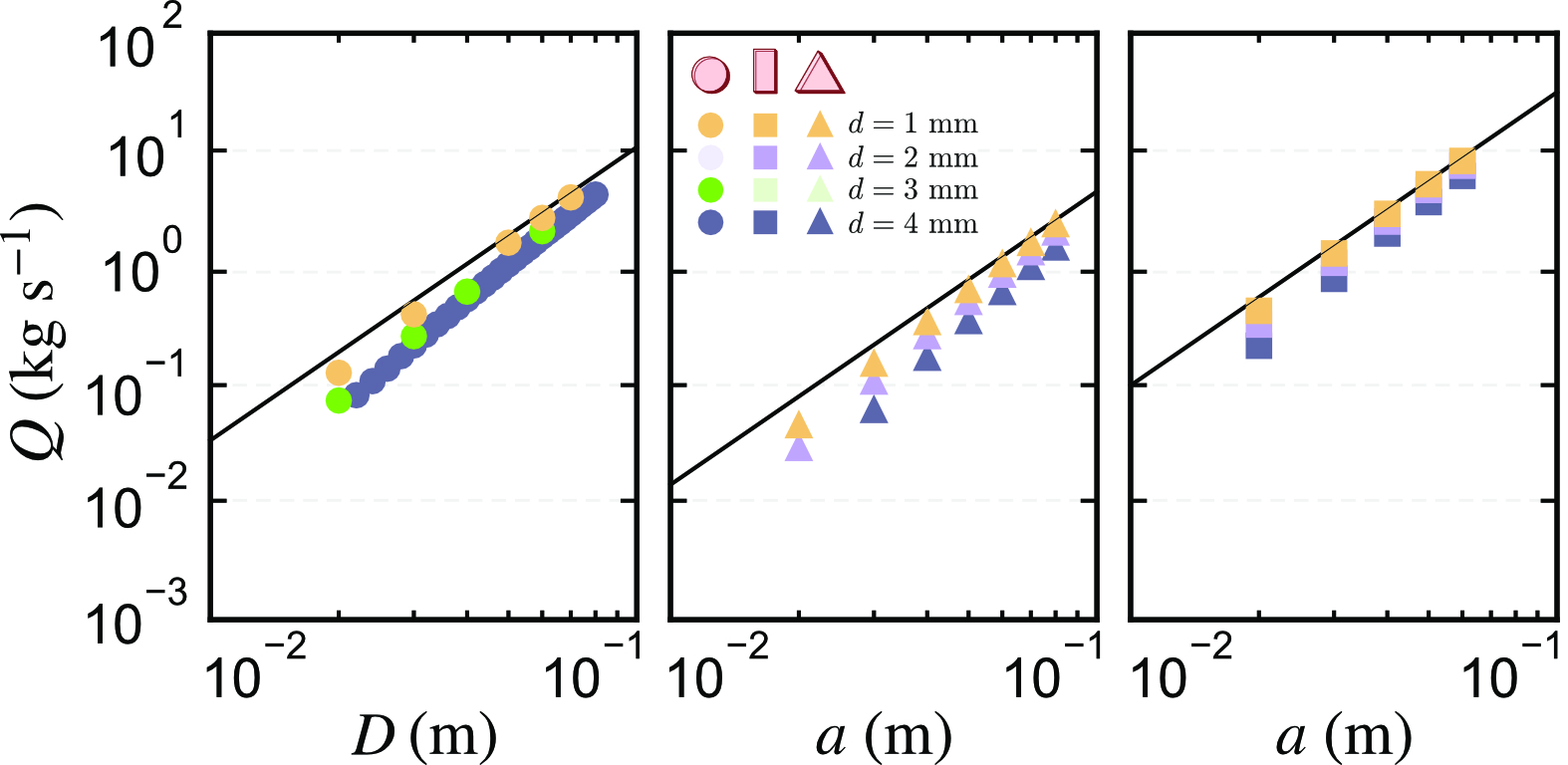}
\caption{Steady mass flux measured from simulations for (a) circular, (b) triangular, and (c) rectangular apertures. Marker colors indicate particle diameter.
The black lines show the free-fall reference $Q_{\rm ff}=A\rho_g\phi_{\rm ff}\sqrt{gD_h}$, with $D_h=4A/P$. Circular-aperture data are from
Ref.~\cite{sharma2026gravity}.}
\label{fig:SFig1_simResults}
\end{figure}

The results of the simulations are shown in Fig.~\ref{fig:SFig1_simResults}, alongside the simulation with circular apertures from our previous study \cite{sharma2026gravity}. The simulated fluxes show trends similar to the
experimental measurements presented in Fig.~2 of the main article.

\begin{table}
\centering
\begin{tabular}{||c|c|c||}
\hline \hline
\textbf{Shape} & \textbf{$a$ [mm]} & \textbf{$D_h$ [mm]} \\
\hline \hline
$\Delta$ & 20 & 11.55 \\
$\Delta$ & 30 & 17.32 \\
$\Delta$ & 40 & 23.09 \\
$\Delta$ & 50 & 28.87 \\
$\Delta$ & 60 & 34.64 \\
$\Delta$ & 70 & 40.41 \\
$\Delta$ & 80 & 46.19 \\
\hline \hline
$\square$ & 20 & 26.67 \\
$\square$ & 30 & 40.00 \\
$\square$ & 40 & 53.33 \\
$\square$ & 50 & 66.67 \\
$\square$ & 60 & 80.00 \\
\hline \hline
\end{tabular}
\caption{Details of apertures used in the simulations.}
\label{tab:aperture_details_simulations}
\end{table}

%%%%%%%%%%%%%%%%%%%%%% Eulerian field construction %%%%%%%%%%%%%%%%%%%

\subsection{Eulerian fields}
During discharge, the positions $\mathbf{x}_p(t) = (x_p, y_p, z_p)$ and velocities $\mathbf{u}_p(t) = (u_{x,p}, u_{y,p}, u_{z,p})$ of every particle $p$ are recorded at $10 \, \mathrm{Hz}$, yielding snapshots at discrete times $t_n$. From these Lagrangian data, we construct the time-averaged occupancy field $\phi(x,y)$ and the time-averaged vertical velocity field $u_z(x,y)$ at the aperture ($z=0$), using the procedure described below. In practice, all fields defined below are evaluated on a uniform Cartesian grid covering the aperture, with spacing $\Delta x = \Delta y = d/50$ and sampling points $\mathbf{x}_{ij} = (x_i, y_j)$ located at the cell centers. Although evaluated on this grid, the fields are written here and in the main text in terms of the continuous variables $(x, y)$ for notational clarity and consistency with the analytical framework.

Building on the three-dimensional field description of our previous study \cite{sharma2026gravity}, we now focus on two-dimensional fields at the aperture. For each particle $p$, the indicator function $I_p(x, y, t)$ takes the value of unity if $(x, y)$ lies within the particle cross-section in the plane $z = 0$, and zero otherwise. The instantaneous occupancy field is then
\begin{equation}
    I(x, y, t) = \sum_p I_p(x, y, t).
    \label{eq:instant_occupy}
\end{equation}
Time averages are taken over the interval $[T_1, T_2]$, defined as in the experiments by the discharge mass $M(t)$, with $M(T_1) = 0.5 \, \mathrm{kg}$ and $M(T_2) = 2.5 \, \mathrm{kg}$.
\begin{equation}
    \phi(x, y) = \frac{1}{T_2 - T_1}\int_{T_1}^{T_2} I(x, y, t)\,\mathrm{d}t \approx \frac{1}{N_T}\sum_{n=1}^{N_T} I(x, y, t_n),
    \label{supp:eq:phi_avg}
\end{equation}
where the discrete approximation sums over the $N_T$ snapshots with $t_n\in[T_1, T_2]$.

The velocity field is constructed from the same indicator functions by first computing the instantaneous occupancy-weighted vertical velocity
\begin{equation}
    j_z(x, y, t) = \sum_p I_p(x, y, t)\, u_{z,p}(t).
\end{equation}
The local time-averaged vertical velocity is then obtained by normalizing the accumulated velocity sum by the accumulated occupancy over the averaging interval,
\begin{equation}
    u_z(x, y) \approx \frac{1}{N_T\,\phi(x, y)}\sum_{n=1}^{N_T} j_z(x, y, t_n),
\end{equation}
defined wherever $\phi > 0$. This definition gives an occupancy-weighted mean velocity and leaves the velocity undefined at sampling points that remain unoccupied throughout the averaging interval.

%%%%%%%%%%%%%%%%%%%%%% Discrete integrals %%%%%%%%%%%%%%%%%%%%
\subsection{Aperture-integrated fields}
Integrated quantities such as the mass flux $Q$ are obtained by summing the discrete fields over the grid points whose cell centers lie within the aperture $\mathcal{A}$, given by the index set
\begin{equation}
    \mathcal{I} = \left\{ (i,j) \;\middle|\; (x_i, y_j) \in \mathcal{A} \right\}.
\end{equation}
For a field $f(x, y)$, the average over the aperture is approximated as
\begin{equation}
    \langle f\rangle_A = \frac{1}{A}\iint_{\mathcal{A}} f(x, y)\,\mathrm{d}x\,\mathrm{d}y \approx \frac{\Delta x^2}{A} \sum_{(i,j)\in\mathcal{I}} f(x_i, y_j),
\end{equation}
consistent with the definition in the main text. The time-averaged mass flux is then computed directly from the instantaneous occupancy-weighted velocity as
\begin{equation}
    Q = A \rho_\mathrm{g}\langle \phi\, u_z \rangle_A \approx \frac{\rho_\mathrm{g}\,\Delta x^2}{N_T} \sum_{n=1}^{N_T}\sum_{(i,j)\in\mathcal{I}} j_z(x_i, y_j, t_n).
\end{equation}

%%%%%%%%%%%%%%%%%%%%%% Details on the packing model %%%%%%%%%%%%%%%%%%%%

\section{Details on the packing model}

For a regular $n$-gon (including the circle as the $n \to \infty$ limit) with inradius $r_{\rm in}$, offsetting every edge inward by a distance $s$ produces a smaller polygon with the same number of sides and the same interior angles - each apothem is reduced by exactly $s$, since the offset is measured perpendicular to every edge simultaneously. The resulting shape is therefore similar to the original, with inradius $r_{\rm in} - s$. Since the area of a family of similar figures scales as the square of any single linear dimension,
\begin{equation}
    \frac{A(s)}{A(0)} = \left[ \frac{(r_{\rm in} - s)}{r_{\rm in}} \right]^2
\end{equation}
Here, $r_{\rm in}=s_{\max}$ is fixed by the aperture geometry. Expanding this expression for small $s$ gives $A(s)=A-2As/s_{\max}+\mathcal{O}(s^2)$. Comparison with $A(s)=A-Ps+\mathcal{O}(s^2)$ therefore yields
\begin{equation}
    s_{\max}=\frac{2A}{P}=\frac{D_h}{2}.
\end{equation}
This relation applies to circles and regular polygons. For the $2{:}1$ rectangular apertures, the inward offsets are not geometrically similar to the original opening.
Instead,
\begin{equation}
    A(s)=(a-2s)(2a-2s),
    \qquad 0\leq s\leq a/2,
\end{equation}
with $s_{\max}=a/2$.

\begin{figure*}
\centering
\includegraphics[width =0.8\textwidth]{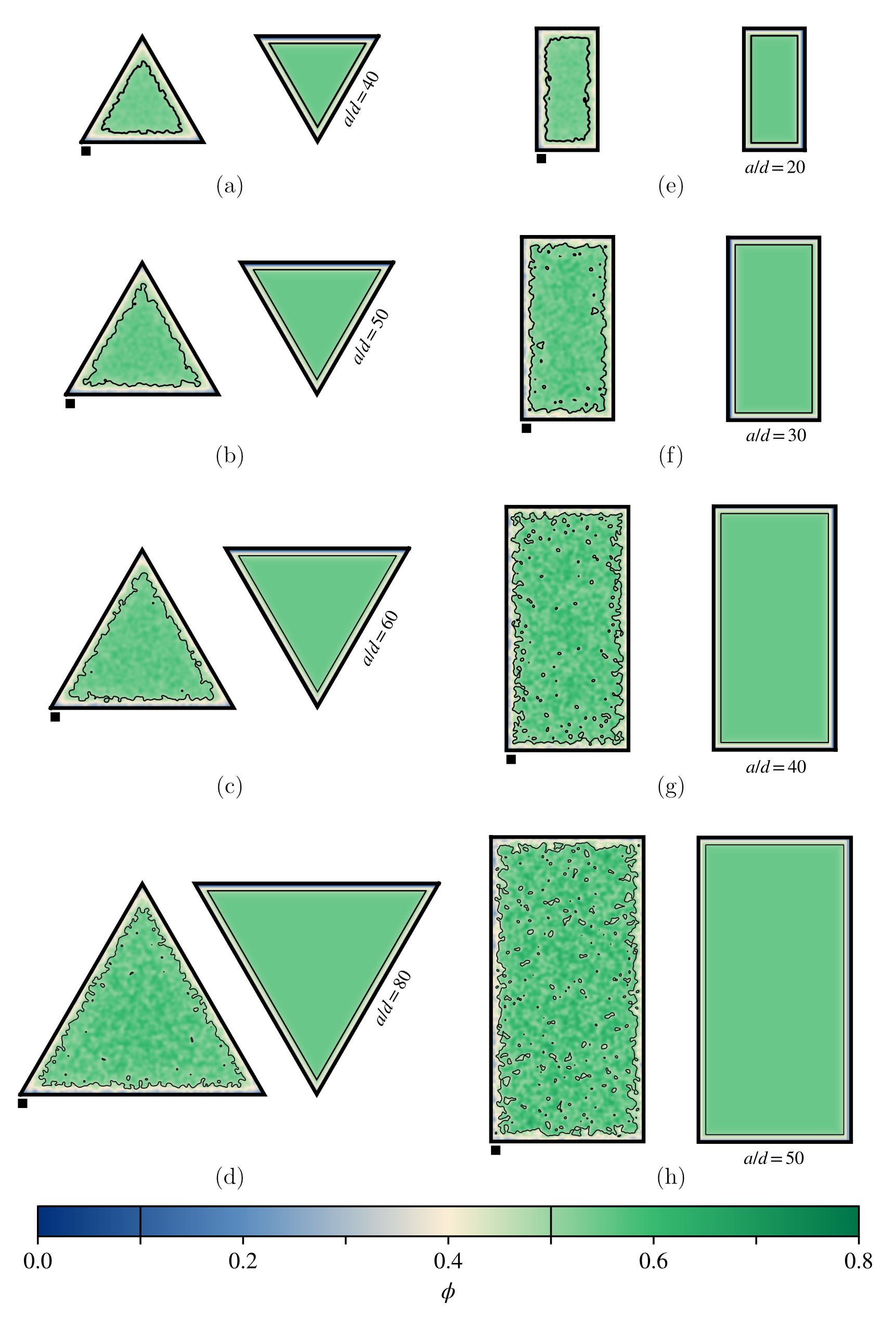}
\caption{Measured (left) and predicted (right) packing fields at the aperture for triangular [(a)--(d)] and rectangular [(e)--(h)] openings. The panels indicate the aperture sizes $a/d$. Predictions use $\phi(s)=\phi_{\rm ff}[1-\exp(-s/\ell)]$, where $s$ is the distance to the nearest aperture boundary, with $\ell=d$. The contours indicate $\phi=0.50$.
}
\label{fig:SFig2PackingPredictions}
\end{figure*}

\end{document}